\documentclass[12pt,preprint,authoryear,round]{imsart}
\RequirePackage[colorlinks,citecolor=blue,urlcolor=blue]{hyperref}
\usepackage{amssymb,amscd,amsthm, verbatim,amsmath,color,fancyhdr, mathrsfs}
\usepackage{graphicx,tabularx,booktabs}
\usepackage{turnstile}
\usepackage{fancyhdr}
\usepackage{caption,subcaption}
\usepackage{url}
\usepackage{color,verbatim,enumerate}
\usepackage[normalem]{ulem}
\usepackage{marginnote} 
\usepackage{multirow}
\usepackage[letterpaper, left=2.5cm, right=2.5cm, top=2.5cm,
bottom=2.5cm,dvips]{geometry}
\usepackage{tikz}

\usepackage[ruled, 
lined, 
commentsnumbered]{algorithm2e}

\theoremstyle{plain} 
\newtheorem{theorem}{Theorem}[section]

\theoremstyle{definition}
\newtheorem{defn}[theorem]{Definition}

\newtheorem{example}[theorem]{Example}

\newcommand{\R}{\mathbb{R}}

\begin{document}
		\begin{frontmatter}
			\title{Algebraic statistics, tables, and networks: \\The Fienberg advantage} 
			\runtitle{Algebraic statistics and networks: the Fienberg advantage}
			
			\author{Elizabeth Gross, Vishesh Karwa, and  Sonja Petrovi\'c}\footnote{ S.P. is at Illinois Institute of Technology, sonja.petrovic@iit.edu; E.G. is at the University of Hawai`i at M\={a}noa, egross@hawaii.edu; V.K. is at Temple University, vishesh@temple.edu. E.G. is supported by NSF DMS-1620109. 
			}

			\runauthor{Gross, Karwa, Petrovi\'c}

\begin{abstract}
Stephen Fienberg's affinity for contingency table problems and reinterpreting models with a fresh look gave rise to a new approach for hypothesis testing of network models that are linear exponential families. We outline his vision and  influence in this fundamental problem, as well as generalizations to multigraphs and hypergraphs. 

\emph{\textcolor{blue}{
}}

\end{abstract}

	\end{frontmatter}	

\section{Introduction}

Stephen Fienberg's early work on contingency tables \cite{BFH:DiscreteMultivariate:1975}
relies on using intrinsic model geometry to  understand the behavior of estimation algorithms,  asymptotics, and model complexity.  For example, in  \cite{FienbergIterative}, Fienberg gives a geometric proof of the convergence of the iterative proportional fitting algorithm for tables with positive entries.  The result in \cite{FienbergIterative} relies on his study of the geometry of $r \times c$ tables in \cite{FienbergGeometry} and his and Gilbert's geometric study of $2 \times 2$ tables \cite{FienbergGilbertGeometry}.  This approach to understanding models would eventually fit within the field of algebraic statistics, a general research direction that would take hold in the 2000s, over 25 years after the publishing of \cite{FienbergIterative} and the 1974 edition of Bishop, Fienberg, and Holland's book \cite{BFH:DiscreteMultivariate:1975}, whose cover displayed the independence model for $2 \times 2$ tables as an algebraic surface. 

The term `algebraic statistics' was coined in 2001~\cite{PistoneWynnEva} and generally refers to the use of broader algebraic---non-linear---and geometric---non-convex--- tools in statistics.   While the use of algebra and geometry had been long present in statistics, before the 2000s, linear algebra and convex geometry were the main tools used consistently. The field of algebraic statistics is now a   branch of mathematical statistics that relies on insights from  computational algebraic geometry, combinatorial geometry, and  commutative algebra to improve statistical inference.   As algebraic statistics matured and caught the attention of many researchers,  
Fienberg and hist students and collaborators reformulated several fundamental statistical problems, e.g. existence of maximum likelihood estimators and ensuring data privacy, into the language of polyhedral and algebraic geometry. 
Today Fienberg's intuition and influence remain central to one of the principal applications in algebraic statistics: testing goodness of fit of log-linear models for discrete data.  Within the last decade or so, much of his work in this area focused on log-linear \emph{network} models.  In this regard, Fienberg defined new models, explained how to represent relational data as contingency tables in order to apply tools from categorical data analysis, and addressed the problems of estimation, model fit, and model selection. 
This paper presents a brief overview of this line of work heavily influenced by Fienberg's vision, which continues to inspire us.


\section{Geometry and algebra of log-linear models}

Let us recall the basics and fix notation. 
Let $\mathcal I = [d_1] \times \cdots \times [d_k]$ be a finite set that indexes  \emph{cells} in a contingency table $u\in\mathbb Z_{\geq0}^{d_1\times\cdots\times d_k}$. 
 The $(i_1,\dots,i_k)$-cell counts the number of  occurrences of the event $\{X_1=i_1, \dots, X_k=i_k\}$ for $k$ categorical random variables with $X_i$ taking values on a finite set $[d_i]:=\{1, \ldots, d_i\}$. 
Log-linear models are probability distributions on the discrete set $\mathcal I$ determined by marginals of the table $u$; since marginalization is a linear map, it can be presented as matrix multiplication.  
Specifically, a log-linear model for $\mathcal I$ is a linear exponential family defined by an $m \times |\mathcal I|$ matrix $A$, called the \emph{design matrix}, taking the following  form: 
\begin{equation}\label{eq:loglinmodel}
	 P_\theta(U=u) = \exp\{\left< Au,\theta\right>-\psi(\theta\}, 
\end{equation}
where $\theta\in\mathbb R^{m}$ is the vector of model parameters and $\psi(\theta)$  the normalizing constant. 
Note that specifying the matrix $A$ completely specifies the contingency table model for $X_1,\dots,X_k$, as it determines the vector of minimal sufficient statistics $Au$ for the linear exponential family in \eqref{eq:loglinmodel}. 
As is customary in algebraic statistics, we will denote the  model \eqref{eq:loglinmodel} by $\mathcal M_A$. 

Let us consider one of Fienberg's early favorite examples: the model of independence of two categorical random variables $X_1$ and $X_2$. Here, $A$ is a $(d_1+d_2)\times d_1d_2$ matrix of the following form, where the first $d_1$ rows each have $d_2$ ones and the last $d_2$ rows contain $d_1$ copies of the $d_2 \times d_2$ identity matrix:

\[
A = \left[\begin{array}{cccc|cccc|c|cccc}1 & 1 & \cdots & 1 & 0 & 0 & \cdots & 0 & \cdots & 0 & 0 & \cdots & 0 \\0 & 0 & \cdots & 0 & 1 & 1 & \cdots & 1 & \cdots  & 0 & 0 & \cdots & 0 \\\vdots & \vdots & \vdots & \vdots & \vdots & \vdots & \vdots & \vdots & \vdots & \vdots & \vdots & \vdots & \vdots \\0 & 0 & \cdots & 0 & 0 & 0 & \cdots & 0 & \cdots & 1 & 1 & \cdots & 1 \\ \\ \hline \\ 1 & 0 & \cdots & 0 & 1 & 0 & \cdots & 0 & \cdots & 1 & 0 & \cdots & 0 \\0 & 1 & \cdots & 0 & 0 & 1 & \cdots & 0 & \cdots & 0 & 1 & \cdots & 0 \\\vdots & \vdots & \ddots & \vdots & \vdots & \vdots & \ddots & \vdots & \vdots & \vdots & \vdots & \ddots & \vdots \\0 & 0 & \cdots & 1 & 0 & 0 & \cdots & 1 & \cdots & 0 & 0 &\cdots & 1\end{array}\right].
\]

The sufficient statistic for $\mathcal M_A$ is the vector of marginal counts (that is, table row and column sums).  For a contingency table $u$, these counts  are computed as:
\[
A	 \left[\begin{matrix} 
	 	u_{11}\\\vdots\\u_{d_1d_2}\\
	  \end{matrix}\right]
	  = 
	 \left[\begin{matrix} 
	 	u_{1+}&\dots &u_{+d_2}
	   \end{matrix}\right]. 
\]
In \cite{FienbergGeometry}, Fienberg describes the geometry of $\mathcal M_A$ in detail, describing the model of independence as the intersection of the \emph{manifold of independence} with the probability simplex. In algebraic geometry, the manifold of independence is a Segre variety, a categorical product, which Fienberg describes explicitly by detailing the linear spaces corresponding to the product fibers.  In addition, the defining  equations of the Segre variety corresponding to the independence model are stated in \cite{FienbergGeometry} in statistical terms (see Section 4 of \cite{FienbergGeometry}).  These equations, which are polynomial equations in indeterminates that represent joint cell probabilities, are a key ingredient to assessing model fit.

 Indeed, assessing model fit for log-linear models, and consequently, log-linear network models, is possible due to a fundamental result in algebraic statistics that establishes a connection between model-defining polynomials  and sampling from the conditional distributions of log-linear models.  The model-defining polynomials of interest are generating sets of polynomial ideals called toric ideals.  The essential component, which binds together the statistical and algebraic, is the vector of (minimal) sufficient statistics for the log-linear exponential family, the vector $Au$ in the definition above. 
 
 One way to perform goodness-of-fit testing for log-linear models, especially in sparse settings such as networks, is to perform Fisher's exact test.  In many cases, however, it is infeasible to compute the  exact conditional $p$-value, thus it is estimated using a Markov chain Monte Carlo (MCMC) method. The \emph{exact conditional $p$-value} of a contingency table $u$ is the probability of a data table being more extreme (less expected) than $u$, conditional on the observed values of the sufficient statistics.  The set of all tables with the same sufficient statistics as $u$ is called \emph{the fiber of $u$ under the model $\mathcal M_A$} and is defined as follows:
$$\mathcal F_A(u) := \{v\in\mathbb Z_{\geq0}^{d_1\times\ldots\times d_k}: Au=Av\}.$$ 
The naming of the reference set $\mathcal F_A(u)$ is derived from algebraic geometry: 
a fiber of a point of the linear map defined by $A$ is its preimage under that map;  in this case, we are considering the set of non-negative integer points in the preimage of the sufficient statistics $Au$.    
 In order to perform the MCMC method to estimate the exact conditional $p$-value, a set of moves must be given, and this set of moves must connect all elements in the fiber $\mathcal F_A(u)$ so that the conditional distribution on the fiber can be sampled properly.  Such a set of moves is called a Markov basis.

\begin{defn}
 A \emph{Markov basis} of the model $\mathcal M_A$ is a set of tables $\mathcal B:= \{b_1,\dots,b_n\}\subset \mathbb Z^{d_1\times\ldots\times d_k}$  for which 
\[
	A b_i =0 
\] 
and such that 
for any contingency table $u\in\mathbb Z_{\geq0}^{d_1\times\ldots\times d_k}$ and 
for any $v\in\mathcal F_A(u)$, there exist $b_{i_1},\dots,b_{i_N}\in \mathcal B$ that can be used to reach $v$ from $u$: 
\[
	u + b_{i_1} + \ldots + b_{i_N} = v
\]
while remaining in the fiber at each step:  
\[
	u+\sum_{j=0}^N b_{i_j} \in \mathcal F_A(u) \mbox{, for $j=1\dots N$}.
\]
\end{defn}
Note that the last requirement simply means that each move needs to preserve non-negativity of cells. 
As an example, let us consider the independence model with $N =2$, $d_1 = 3$, and $d_2 = 3$.  Then the fiber $F_A(u)$ for any $u$ is a collection of $3 \times 3$ tables. Examples of three different Markov moves for the independence model in this setting are

\[ \begin{array}{|c|c|c|}\hline 1 & -1 & 0 \\\hline -1 & 1 & 0 \\\hline 0 & 0 & 0 \\\hline \end{array}, \ \ \begin{array}{|c|c|c|}\hline -1 & 0 & 1 \\\hline 0 & 0 & 0 \\\hline 1 & 0 & -1 \\\hline \end{array},  \ \ \begin{array}{|c|c|c|}\hline 0 & 0 & 0 \\\hline 0 & 1 & -1 \\\hline 0 & -1 & 1 \\\hline \end{array}.\]

It is hard to check \emph{a priori} whether a given set of moves does in fact form a Markov basis for the model.  However, the following foundational result from algebraic statistics, allows one to compute a Markov basis by computing a generating set of a polynomial ideal.

\begin{theorem}[\cite{DS98}]\label{thm:FTMB}
A  set of vectors $\mathcal B=\{b_1,\dots,b_n\}$ is a Markov basis of the log-linear model $\mathcal M_A$ \emph{if and only if} the corresponding set of binomials 
$
	\{ x^{b_i^{+}} - x^{b_i^{+}} \}_{i=1,\dots,n}
$
generates the toric ideal $I_A:=( x^u-x^v : u-v\in\ker_{\mathbb Z}A )$. 
\end{theorem}

Considering again the independence model with $N =2$, $d_1 = 3$, and $d_2 = 3$, the binomials associated to the three tables above are:
\[ x_{11}x_{22} - x_{12}x_{21}, \ \ \ x_{13}x_{31} - x_{11}x_{33}, \ \ \ x_{22}x_{33} - x_{23}x_{32}. \]
One can check that these three polynomials are not enough to generate the ideal $I_A$, and thus more moves are needed for a Markov basis.

Theorem \ref{thm:FTMB} is a powerful result the connects categorical data analysis to algebra. By connecting network analysis to categorical data analysis, Fienberg was able to use the full force of this theorem for testing model fit of statistical network models.

\section{Log-linear ERGMs and goodness-of-fit testing}

As stated in the editorial piece \cite{JAS-editorial-2019}, Fienberg took joy in rediscovering old concepts from new points of view that gave them new interpretations and wider applicability; this was evident not only from his research articles and conference presentations, but various interviews, see, for example, \cite{InteviewWithSteve}. We follow his lead in the way we define \emph{log-linear network models}. 

Generally, a statistical network model is a collection of probability distributions over $\mathcal G_n$, the set of all (un)directed graphs on $n$ vertices. The Fienberg approach to the analysis of statistical network models, dating back to the late '70's and early '80's, relies on explicitly making the connection to categorical data analysis by viewing graphs as contingency tables.  
  For example, in \cite{FienbergWasserman1981categorical}, Fienberg and Wasserman view a directed graph with $n$ vertices as a $n \times n\times2\times2$ table $Y$ where $Y_{ij00}=1$ if there is no edge between vertex $i$ and $j$, $Y_{ij11}=1$ if there is a reciprocated edge between $i$ and $j$, $Y_{ij10}=1$ if there is a non-reciprocated edge from $i$ to $j$, and $Y_{ij01}=1$ if there is a non-reciprocated edge from $j$ to $i$, and all entries are $0$ otherwise. Using this $n \times n\times2\times2$ table, Fienberg and Wasserman then describe nine variants of a simple statistical network model, called the $p_1$ model \cite{HL81}, in terms of table marginals and show how these models can be fit using a version of iterative proportional scaling for multidimensional contingency tables.  In addition, they also develop a variant of the $p_1$ model for $K$ subgroups determined by nodal attributes, by collapsing the $n \times n\times2\times2$ into a $K \times K\times2\times2$ table;  a precursor to the directed stochastic blockmodels. 

The $p_1$ model and its variants described by Fienberg and Wasserman in \cite{FW81} are examples of log-linear ERGMs. Log-linear ERGMs are exponential family random graph models with a log-linear interpretation. Another example of log-linear models are stochastic blockmodels, which are given a contingency representation in \cite{FienbergMeyerWasserman1985block}. Following the contingency table framework of the Fienberg approach, to define a log-linear ERGM, one chooses an embedding $\phi:\mathcal G_n \to \mathbb R^{\ell}$ such that for all $G=(V,E)$ we have $\phi(G)=\sum_{e \in E} \phi(e)$, and implicitly uses the embedding $\phi$ to represent $G$ as a vector. For example, for directed graphs, a reasonable embedding would embed $\mathcal G_n$ into $\mathbb R^{ n^2}$ and $G$ would be represented by its vectorized adjacency matrix, while for undirected graphs $\mathbb R^{n \choose 2}$ would work equally well. For directed graphs, a suitable embedding rooted in \cite{FienbergWasserman1981categorical} (see also \cite{FW81}) maps $\mathcal G_n$ into $\mathbb R^{n \times n\times2\times2}$ by representing graphs by their vectorized $n \times n\times2\times2$ Fienberg-Wasserman table or a vectorized table of size ${n \choose 2} \times 2 \times 2$ after removing redundant cells.  These embeddings allows us to refer to graphs as vectors. 

An exponential family random graph model, or an ERGM for short, is a collection of probability distributions on $\mathcal G_n$ that places the following probability on each graph  $G\in\mathcal G_n$: 
\begin{equation}\label{eq:ERGM}
P_{\theta}(G) = Z(\theta) e^{{\theta} \cdot t(G)}, 
\end{equation}
where $G$ is uniquely represented as a vector in $\R^{\ell}$, 
$\theta$ is a row vector of parameters of length $q$, the map $t: \R^{\ell} \to \R^q$ 
computes the sufficient statistics, and $Z(\theta)$ is a normalizing constant.  The image of the sufficient statistic map $t$ is a vector in which each entry is a network statistic used to specify the model, such as edge count, degree of a given vertex,  number of edges in a given block of vertices, etc.  When the sufficient statistic is a linear function on the entries of a natural contingency table representation of the graph, as in degree-based models or stochastic blockmodels, then the sufficient statistic map $t$ can be described with a design matrix $A$ and the model \eqref{eq:ERGM} takes the form of \eqref{eq:loglinmodel}. When this happens, we call the model a log-linear ERGM.

\begin{defn}\label{defn:loglinearERGM}
We call such a model a \emph{log-linear ERGM} if 
the sufficient statistic map $t$ in the ERGM specification~\eqref{eq:ERGM}  is a \emph{linear} map $t: \R^{\ell} \to \R^q$ from the space of graphs to the space of the minimal sufficient statistics of the model.
\end{defn}

Log-linear ERGMs include degree-based models such as the $\beta$-model, models that include effects for reciprocity, such as $p_1$ models, and models for data with categorical nodal attributes, such as stochastic blockmodels.  Since the sufficient statistic $t$ is a linear map, dyadic independence is implied for a log-linear ERGM.  Dyadic independence is another way to say that for each pair of vertices, $i$ and $j$, the edge configuration (e.g., no edge between $i$ and $j$, directed edge from $i$ to $j$, directed edge from $j$ to $i$, bidirected edge between $i$ to $j$) is independent from the edge configuration between any other pair of vertices.  Thus, we can fully specify a log-linear ERGM by specifying the distribution over each set of dyadic configurations.   

\begin{example}[Stochastic blockmodels] \label{eg:blockmodels} Extremely popular in practice, this family of log-linear ERGMs models networks whose nodes are partitioned into groups--blocks-- according to some nodal attributes. For a directed network, each dyad can be in one of four states represented as follows: $(0,0)$ represents no edge, $(1,0)$ an edge from $i$ to $j$, $(0,1)$ an edge from $j$ to $i$, and $(1,1)$  a bidirected edge.  Note that if the network is undirected, the model simply collapses to having only two dyadic states: (0,0) and (1,1).  Denote by $p_{ijkl}$ the probability of the dyad  $(i,j)$ to be in state $(k,l)$. 

Edge formation is governed by what Fienberg and Wasserman call choice parameters, denoted by $\delta^{rs}$, and    reciprocity effects $\rho^{rs}$. These parameters are defined on the level of blocks. In addition, Fienberg liked the use of an additional set of parameters $\lambda_{ij}$ for normalization: ensuring that each dyad is observed in only one state at a time.  Specifically, the model was defined in \cite{FienbergMeyerWasserman1985block}  as follows: 

\begin{align}\label{eq:FMWblockmodel}
	\log p_{ij00} &= \lambda_{ij}  \\
\nonumber	\log p_{ij10} &=  \lambda_{ij} + \delta^{b(i)b(j)}\\
\nonumber		\log p_{ij01} &=  \lambda_{ij} + \delta^{b(j)b(i)}\\
\nonumber		\log p_{ij11}&= \lambda_{ij}+\delta^{b(i)b(j)}+\delta^{b(j)b(i)}+\rho^{b(i)b(j)}, 
\end{align}
where   each node in the graph belongs to one of $K$ blocks, $B_1,\dots,B_K$, and $b(i)$ denotes the  (known) block assignment of vertex $i$.  
 
 There are various special cases of stochastic blockmodels. For example, we can choose $\delta^{rs}=\delta+\alpha^r+\beta^s$ and $\rho^{rs}=\rho$, as in \cite[Equation (2.10)]{FienbergMeyerWasserman1985block}. Then the model is the following special case: 
 
 \begin{align}\label{eq:FMWblockmodel}
	\log p_{ij00} &= \lambda_{ij}  \\
\nonumber	\log p_{ij10} &=  \lambda_{ij} + \delta+\alpha^{b(i)}+\beta^{b(j)}\\
\nonumber		\log p_{ij01} &=  \lambda_{ij} + \delta+\alpha^{b(j)}+\beta^{b(i)}\\
\nonumber		\log p_{ij11}&= \lambda_{ij}+2\delta+\alpha^{b(i)}+\alpha^{b(j)}+\beta^{b(j)}+\beta^{b(i)}+\rho, 
\end{align}
 
\noindent In this setting, the sufficient statistics counted by the map $t$ are the number of configurations for each dyad, the total number of edges, block in-degrees, block out-degrees, and the total number of reciprocated edges in the network. Here, the in-degree of block $B_j$ (the number of edges that enter the block) is computed by adding in-degrees of all the nodes in the block,  $d_{B_j}^{in}=\sum_{i\in B_j} d_i^{in}$. The out-degree is defined similarly. 

Let's consider the space of directed graphs on $n=3$ vertices $V =\{1,2,3\}$ with block structure $B_1=\{1,2\}$, $B_2 = \{3\}$, the design matrix $A$ defining the linear map $t$ would be as follows

\[ A = \left[\begin{array}{cccccccccccc}1 & 1 & 1 & 1 & 0 & 0 & 0 & 0 & 0 & 0 & 0 & 0 \\0 & 0 & 0 & 0 & 1 & 1 & 1 & 1 & 0 & 0 & 0 & 0 \\0 & 0 & 0 & 0 & 0 & 0 & 0 & 0 & 1 & 1 & 1 & 1 \\0 & 1 & 1 & 2 & 0 & 1 & 1 & 2 & 0 & 1 & 1 & 2 \\0 & 1 & 1 & 2 & 0 & 1 & 0 & 1 & 0 & 1 & 0 & 1 \\0 & 0 & 0 & 0 & 0 & 0 & 1 & 1 & 0 & 0 & 1 & 1 \\0 & 1 & 1 & 2 & 0 & 0 & 1 & 1 & 0 & 0 & 1 & 1 \\0 & 0 & 0 & 0 & 0 & 1 & 0 & 1 & 0 & 1 & 0 & 1 \\0 & 0 & 0 & 1 & 0 & 0 & 0 & 1 & 0 & 0 & 0 & 1\end{array}\right]. \]

\noindent Let $G$ be represented as a vector of length $12$, where the first four entries correspond to the four possible dyadic configurations between vertices $1$ and $2$, the second four correspond to the four possible dyadic configurations between vertices $1$ and $3$, and the third four correspond to the four possible dyadic configurations between vertices $2$ and $3$. Then the first three rows of $A$ count the number of configurations for each dyad (for simple graphs this count should always be one), the fourth row of $A$ counts the total number of edges in $G$, the fifth and sixth rows count the block in-degrees, the seventh and eighth rows count the block out-degrees, and last row counts the total number of reciprocated edges in the network.
\end{example}

\begin{example}[$p_1$ models] The $p_1$-model for directed graphs was introduced by Holland and Leinhardt \cite{HL81} and extended by Fienberg and Wasserman \cite{FW81}.  It is a model that includes two nodal effects, one for popularity and another for expansiveness, and a reciprocation effect. Following Example \ref{eg:blockmodels}, we denote $p_{ijkl}$ the probability of the dyad $(i,j)$ to be in state $(k,l) \in \{0,1\}^2$.  
The dyadic probabilities for the $p_1$-model are specified as follows:
\begin{align}
	\log p_{ij00} &= \lambda_{ij},  \\
\nonumber	\log p_{ij10} &=  \lambda_{ij} + \alpha_i+\beta_j+\delta,\\
\nonumber		\log p_{ij01} &=  \lambda_{ij} +\alpha_j+\beta_i+\delta,\\
\nonumber		\log p_{ij11}&= \lambda_{ij}+\alpha_i+\alpha_j+\beta_j+\beta_j+2\delta+\rho_{ij}. 
\end{align}
The parameters $\alpha_i$  and $\beta_i$ record the rates at which the node $i$  sends and receives links, while $\rho_{ij}$ controls reciprocation. Note that the model specification includes additional parameters. Namely, there is $\delta$, a density parameter and ${n\choose 2}$ dyadic effects, $\lambda_{ij}$, which are normalizing constants as described in Example \ref{eg:blockmodels}.

The $p_1$ model has three main variants that capture different reciprocation effects: zero reciprocation, constant reciprocation, and dyad-specific reciprocation, also referred to as differential reciprocity. For example, in the constant reciprocation case, $\rho_{ij} = \rho$ for all $i,j$.  The sufficient statistics for the $p_1$-model with constant reciprocation consists of the number of edges, the in-degree sequence, the out-degree sequence, and the number of reciprocated edges.

The design matrix $A$ for several small examples can be found in \cite{PRF:09}. 
\end{example}

\medskip

While Fienberg's work allows for a transfer of technology from the contingency table literature to networks,  the interpretability of models and model equivalence was not always immediate and required additional insight. As noted in \cite{Habe:1981} and reiterated by Fienberg and co-authors in \cite{SteveAleMe-holland}, even simple ERGMs, such as the $p_1$ model, pose fundamental challenges to the practitioner even within the contingency table setting, especially when testing  model goodness of fit. For example, as pointed out by Fienberg and co-authors in \cite{PRF:09}, many network models such as the $p_1$ model are theoretically problematic, since, in these models, the number of parameters depends on the number of vertices. This means that as the population size grows, the model complexity also increases, unlike traditional statistical models, where the complexity is often fixed and independent of the sample size. Another challenge to using existing traditional methods from categorical data analysis in goodness-of-fit testing and model selection is that the data are naturally sparse, making standard asymptotic methods unreliable.  Under such conditions, exact conditional tests are preferred for model selection and goodness-of-fit testing.  However, as mentioned in the previous section, exact conditional tests pose their own difficult problems for networks, mainly since the exact distribution is over a space that is combinatorially large, and in most cases, innumerable. Finally, the contingency tables described by Fienberg and Wasserman are highly redundant and are subject not only to symmetric constraints but also product multinomial constraints, e.g. since each dyad can only be in one of the four possible configurations $Y_{ij00}+Y_{ij10}+Y_{ij01} + Y_{ij00} = 1$ for all $i \neq j$. 

Fienberg was able to provide a work-around to the difficulties posed by exact conditional tests by using Markov bases and algebraic statistics. In 1998, Sturmfels and Diaconis published Theorem \ref{thm:FTMB} \cite{DS98}. Afterwards, the idea of using toric ideals for goodness-of-fit testing for various log-linear models gained traction, and about ten years later, Fienberg, Petrovi\'{c}, and Rinaldo applied Theorem \ref{thm:FTMB} to three of the main variants of the $p_1$ model in \cite{PRF:09}, essentially introducing algebraic statistics to the field of network analysis. In particular, they describe Markov moves for each variant and its corresponding simplified model (the model obtained after forgetting the normalizing parameters).  The work not only provided a breakthrough in goodness-of-fit testing for log-linear ERGMs, but also had an impact in combinatorial commutative algebra.  The toric ideals corresponding the $p_1$ model are connected to toric ideals of graphs, defined in \cite{SVV94} (see also \cite{Vill95} and \cite{OH-multipartite}) and more generally, toric ideals of hypergraphs. Indeed, the results of \cite{PRF:09} provided an applied motivation for the systematic study of toric ideals of hypergraphs  in the field of combinatorial commutative algebra (see e.g. \cite{GrPe}, \cite{HaVanTuylHypergraphs}, \cite{PShypergraphs}, \cite{PTVhypergraphs}). 

\begin{figure}[h]
	\includegraphics[scale=.5]{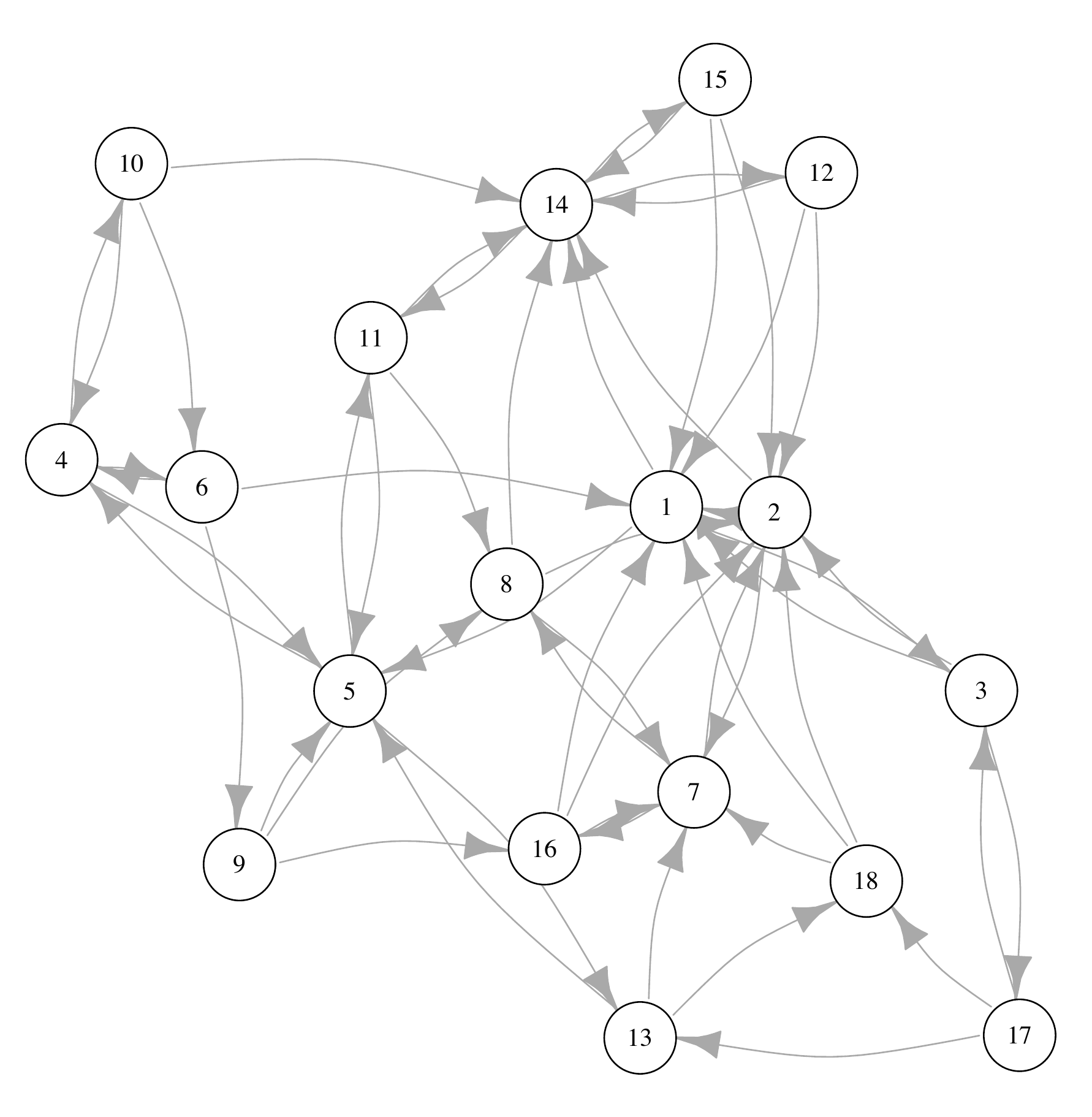}
	\caption{The directed graph representation of Sampson's monastery dataset  \cite{Sampson68}.} 
\label{fig:MonkData}
\end{figure}

\begin{figure}[h]
  \includegraphics[scale=.2]{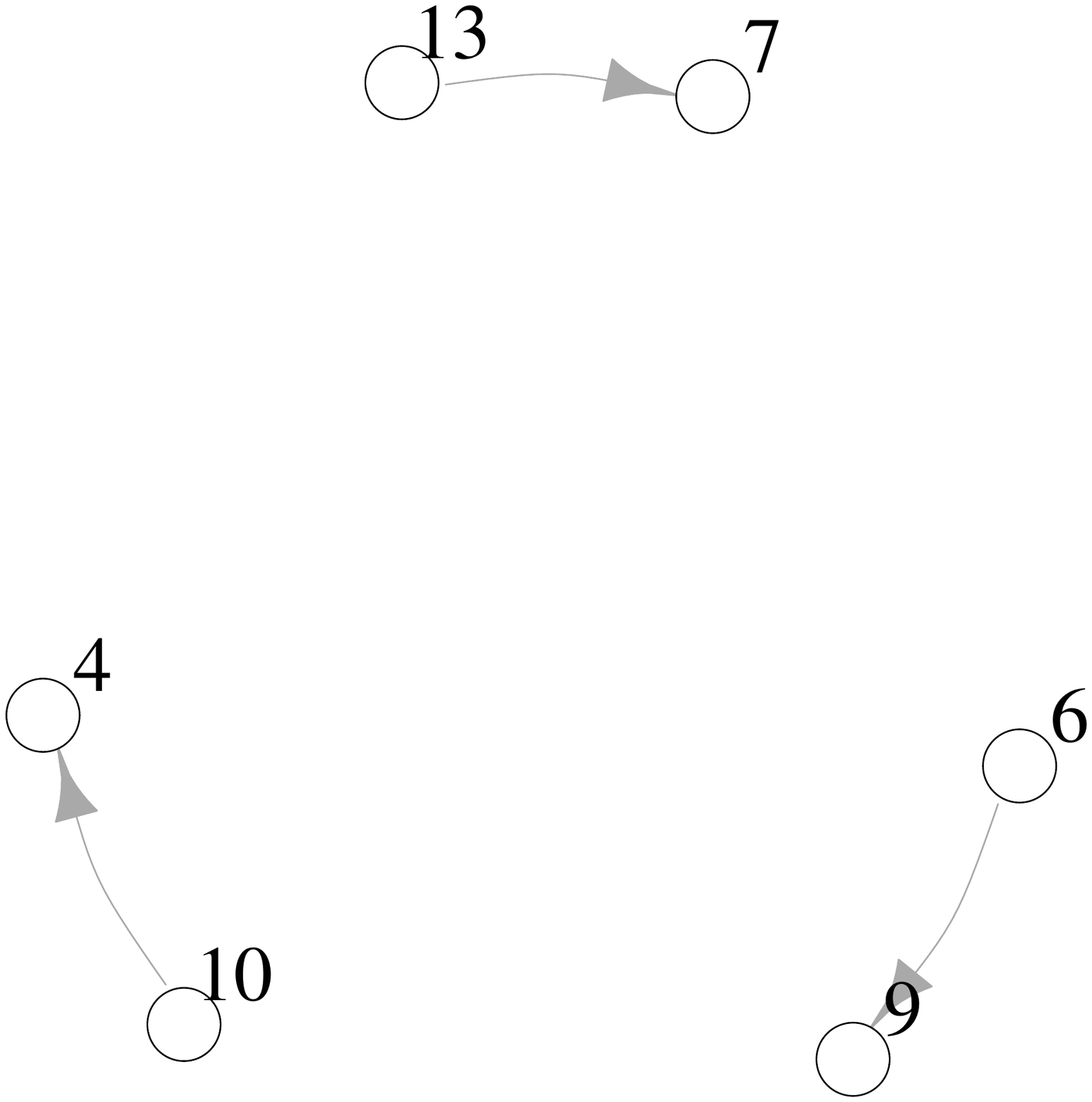} 
 \qquad \qquad \qquad \qquad
 \includegraphics[scale=.2]{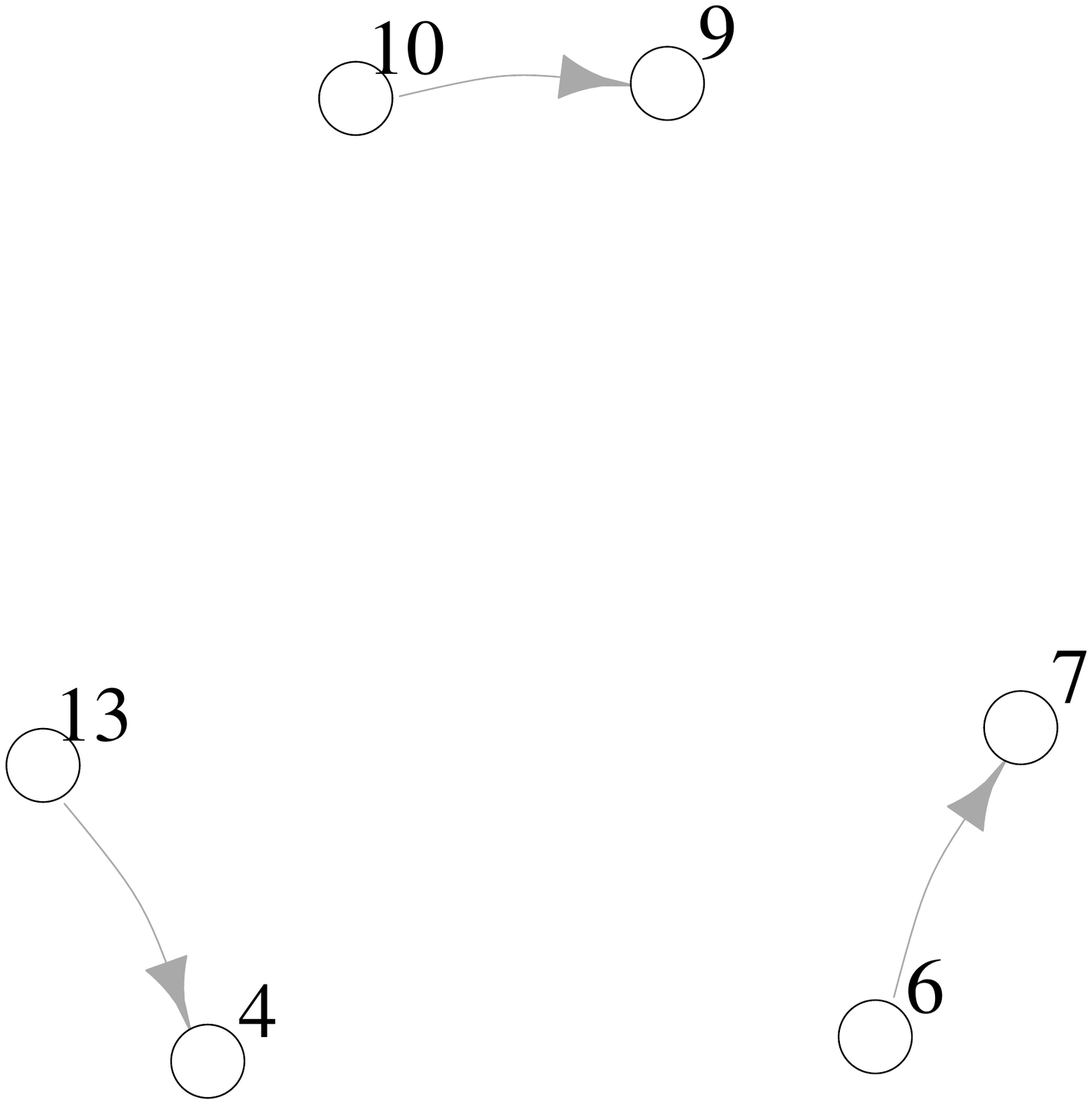}
	\caption{A move from the Markov basis for the $p_1$ model with zero reciprocation. {\bf Left:} Edges to remove. {\bf Right:} Edges to add.  This move can be applied to the network in Figure~\ref{fig:MonkData} as it preserves node in-degrees and out-degrees. Note that edge $4\leftarrow10$ is reciprocated in the data, so after the move is applied, the total number of reciprocated edges is reduced by $1$.} 
\label{fig:MonksMove}
\end{figure}

\begin{figure}[h]
  \includegraphics[scale=.2]{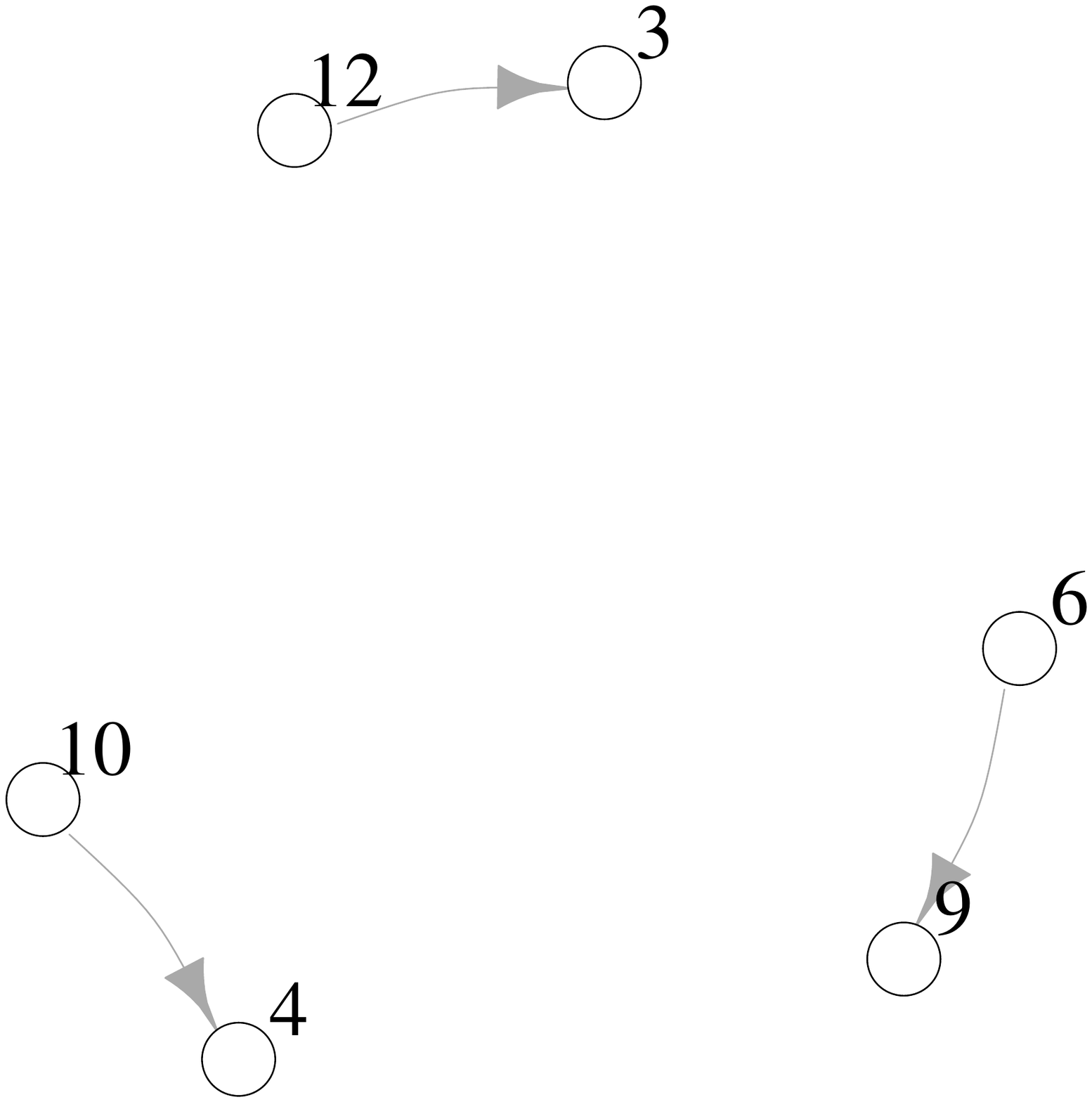} 
  \qquad \qquad \qquad \qquad
 \includegraphics[scale=.2]{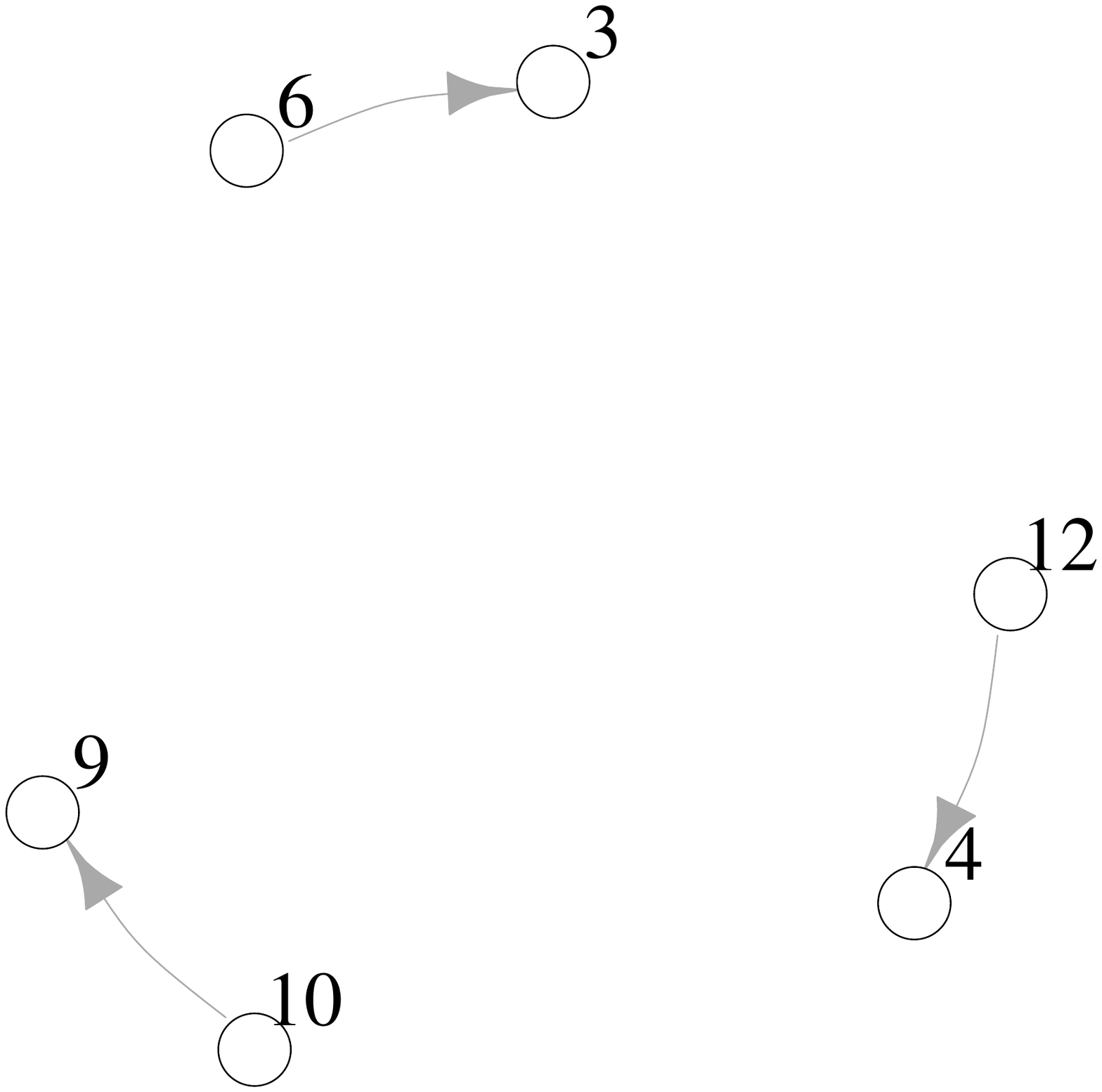}
	\caption{A move from the Markov basis for the $p_1$ model with zero reciprocation. {\bf Left:} Edges to remove. {\bf Right:} Edges to add.  
	 However, this move cannot be applied to the network in Figure~\ref{fig:MonkData} as the dyad $(3,12)$ is observed in the state $(0,0)$ rather than $(1,0)$; that is, the edge  $12\rightarrow3$ is not present, so  it cannot be removed.} 
\label{fig:MonksMoveInapplicable}
\end{figure}

\begin{figure}[h]
   \includegraphics[scale=.2]{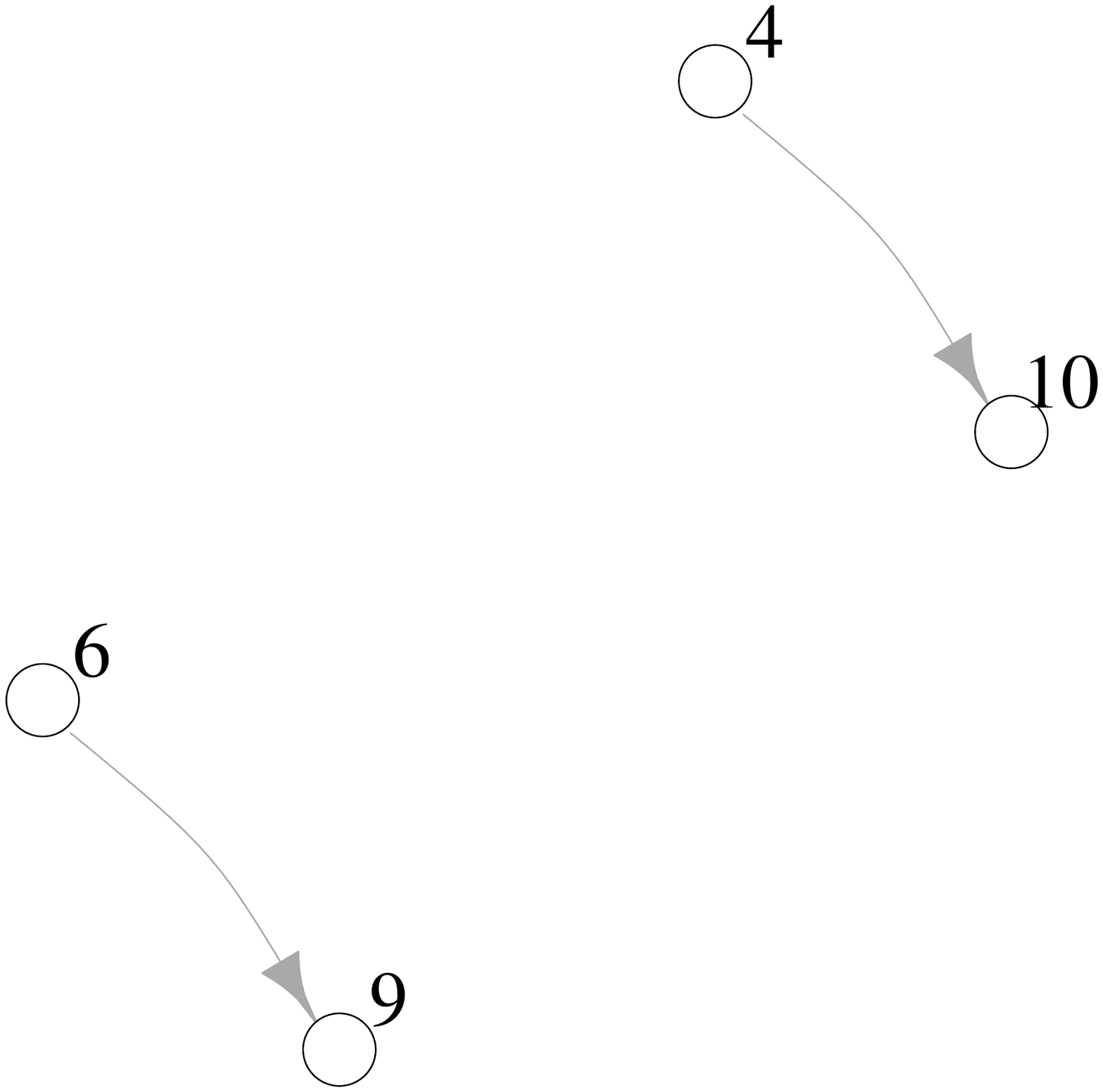} 
   \qquad \qquad \qquad \qquad
 \includegraphics[scale=.2]{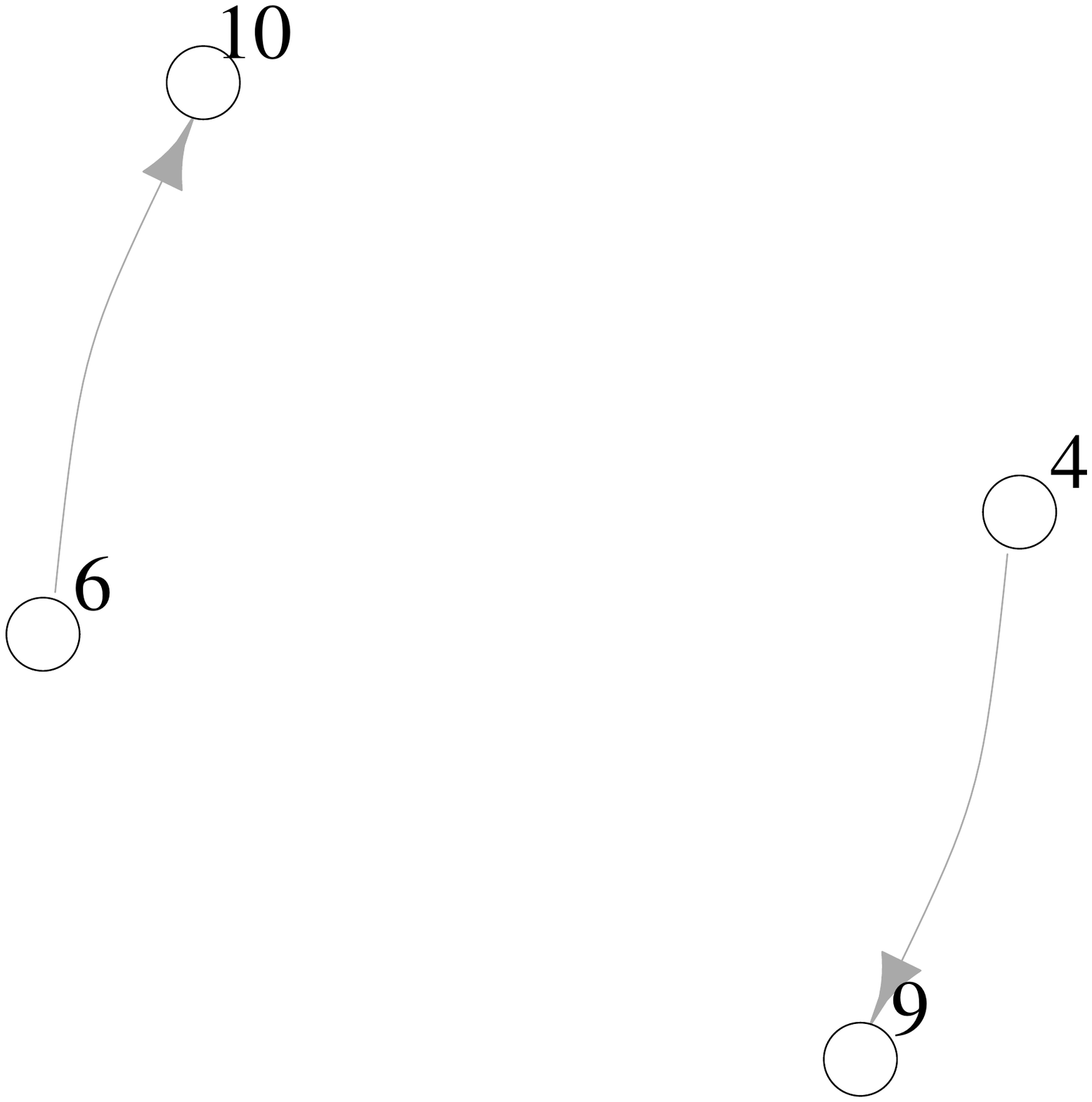}
	\caption{A move from the Markov basis for the $p_1$ model with constant reciprocation. {\bf Left:} Edges to remove. {\bf Right:} Edges to add.   This move can be applied to the network  in Figure~\ref{fig:MonkData}. It preserves the number of reciprocated edges: the dyad $(4,10)$ changes from reciprocated to directed edge, but the dyad $(6,10)$ changes from directed to reciprocated.} 
\label{fig:MonksMoveConstantrecip}
\end{figure}

Before \cite{PRF:09}, 
 Markov bases were always used in the setting where the only constraints on the contingency tables were that every entry needed to be non-negative.  
  However, in the network setting, particularly in the case of a single sociometric relation, cells of the contingency tables are either 0 or 1 and there is only a single observation for each dyad. This was the first time in the Markov bases literature that sampling constraints of this form were directly incorporated in the study of Markov bases (note that related work \cite{HT:10}, and relevant for the problem here,  on connecting tables with 0/1 entries appeared in the same volume).  Fienberg and co-authors were able to effectively handle the network constraints by computing a minimal generating set of this ideal first and then by removing basis elements that violate the condition of one observation per dyad, which results in a product multinomial sampling scheme. 
  Fienberg's idea of adding the normalizing parameters  $\lambda_{ij}$s to the models 
directly enforced the $0/1$  constraint in sampling. In particular, if a move produced by a Markov basis computation is applicable to the observed network, in that it doesn't attempt to remove edges that are not present, then it will follow the sampling constraint in that it will not add an edge where there is one already. 
Examples of applicable and inapplicable moves for the $p_1$ model and the Sampson data depicted in Figure~\ref{fig:MonkData} are shown in Figures~\ref{fig:MonksMove}, \ref{fig:MonksMoveInapplicable}, and \ref{fig:MonksMoveConstantrecip}. 

It should be noted that Fienberg's idea to prune non-applicable moves was novel and paved the way for practical implementation of a goodness-of-fit test for log-linear ERGMs \cite{GPS16}. Indeed, in \cite{DobraEtAl-IMA}, Fienberg and co-authors observed that Markov bases are \emph{data independent}, meaning that they describe all the moves required to guarantee connectedness of any fiber; in other words, Markov bases do not depend on the observed network, only the model.  This observation can help transform otherwise unwieldly sets of Markov moves into smaller and easier to manipulate sets of moves.  For example, without pruning, the naive computation of a Markov basis for the $p_1$ model with constant reciprocation with 4 nodes has 80,610 moves, while the pruned Markov basis consisting of only elements applicable to simple networks and decomposed into essential building blocks, computed in \cite{PRF:09}, has about 10 moves. 

This idea was a starting point of departure from the algebraic status-quo approach, which is traditionally blind to data and as such leads to slow mixing times of the resulting Markov chains. After Fienberg's work in \cite{PRF:09}, the main computational challenge remained open to make the theory useful for network data in practice. To this end, working within the \emph{data dependent paradigm}, \cite{GPS16} developed an algorithm to approximate the exact conditional $p$-value for log-linear ERGMs and implemented the algorithm for the $p_1$ model.  The algorithm approximates the exact conditional $p$-value by using applicable Markov moves generated on an as-needed basis to move around the fiber. At each network in the chain, a goodness-of-fit statistic is computed and compared to the observed network.  This adapted Metropolis-Hastings algorithm is described in detail in \cite{GPS16}.

Fienberg saw great value in small data; thus \cite{GPS16} revisited the Sampson's monastery dataset \cite{Sampson68} (see Figure~\ref{fig:MonkData}) and tested the fit of the $p_1$ model.  The Sampson's monastery dataset, in Fienberg's words, was one of the reasons behind the construction of the Holland-Leinhardt $p_1$ model in the first place. However, the ideas described here do scale, e.g. \cite{KP:AOAS} tests model fit for the $\beta$ and $p_1$ models on coauthorship and citation networks of statisticians \cite{JiJinAOAS} of about 3000 authors and 3000 papers. 
Finally, Fienberg was also an avid supporter of applications of statistics; it was he who suggested to the third author to study the Japanese corporate data set from \emph{The New York Times} back in 2014 from the point of view of the $p_1$ model. As \cite{WhatIsAMarkovBasis} illustrates, the goodness of fit test confirms Japanese Prime Minister's intuition.

\section{Beyond simple graphs} 

The rapid increase of data-collecting mechanisms in  recent decades has resulted in complex forms of network data, including  multivariate  and multi-agent networks. Still, in the growing field of network science, such data are still often represented in the form of a simple graph, mainly because simple random graph models are assumed to be easier to estimate and fit. However,
such simplifications are not necessary with Fienberg's view of networks as contingency tables.  This is because neither multiple observations on a single dyad, which increase cell counts in the table, nor multi-way interactions, which increase table dimensions, present an additional layer of difficulty for estimation or testing model fit. On the contrary, the sampling algorithms based on Markov bases become easier, because the sampling constraint is relaxed. 

One example of this simplification is when experiment data consisting of multiple observations is  summarized as a simple graph by way of thresholding---preserving an edge between two nodes only if it was observed at least a fixed number of times. This happens very often in neuroscience and chemical reaction experiments. It is also often applied to social interactions data such as the coauthorship network in the Figure \ref{fig:hypergraphVSgraph} below.  In the coauthorship network, an edge $(i,j)$ is present in the coauthorship graph if at least 4 joint papers were written by authors $i$ and $j$. Why $4$?  
 This thresholding number of choice seems arbitrary at best (changing it may drastically change the structure of the  graph), is done out of convenience, and in many applications results in significant information loss. 
 
In \cite{FMWTechnicalReport} and \cite{FienbergMeyerWasserman1985block}, Fienberg, Meyer, and Wasserman set up the log-linear framework for multivariate directed graphs.  We can think of a multivariate graph as a multi-layered network.  For example, in the technical paper \cite{FMWTechnicalReport}, Fienberg, Meyer, and Wasserman consider a community of individuals and networks formed by three relations, information, money, and support; these relations are referred to as sociometric generators.  In  \cite{FienbergMeyerWasserman1985block}, the authors develop extensions of \cite{FMWTechnicalReport} to allow for covariates.
Motivated by this, \cite{RPF:11} (see also \cite{RPF:10} for further details) study the \emph{generalized} $\beta$-model for random graphs. They consider the log-linear model for undirected graphs whose sufficient statistics are node degrees, but they allow for the possibility that
each dyad in the network be sampled a different number of times. 
Applying the geometric and combinatorial properties of log-linear models under product-multinomial sampling schemes from \cite{SteveAle12}, they derive necessary and sufficient conditions for MLE existence and discuss its asymptotics. 

\begin{figure}[h]
	\includegraphics[scale=.5]{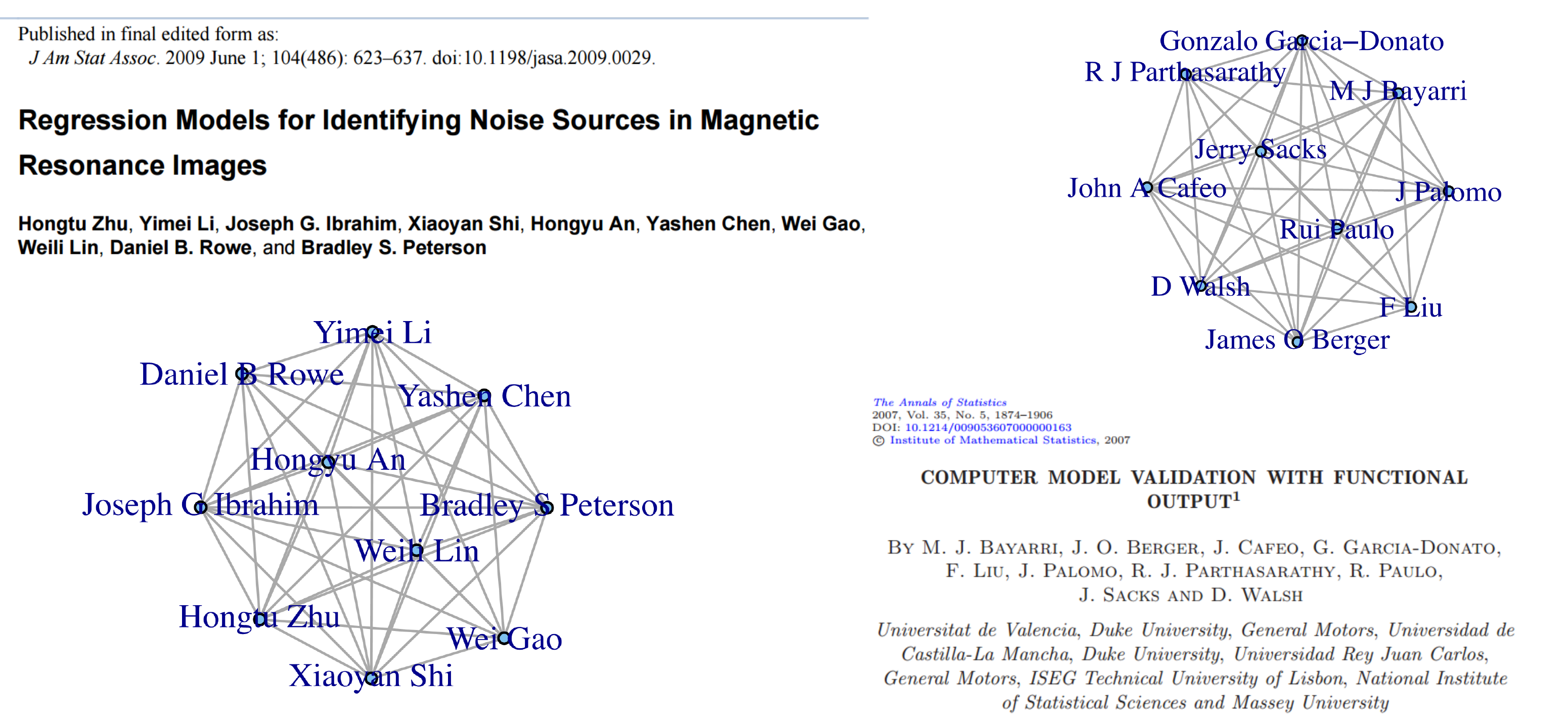}
	\caption{The inner-most clique of each of the two co-authorship graphs studied in \cite{JiJinAOAS}: each corresponds to a many-author paper. Graphs in the figure adapted from \cite{KP:AOAS}.}
\label{fig:CliqueSecret}
\end{figure}

The second example of data simplification is also well illustrated using  coauthorship data: it is common for multiway interactions to be collapsed to their induced pairwise interactions. However, most of the time, capturing the multiway interaction is more realistic and informative.  Figure~\ref{fig:hypergraphVSgraph} shows how the information from  data that naturally comes in form of a hypergraph is obscured when represented by the underlying graph. Indeed, once the first coauthorship network data for statisticians was collected and released in \cite{JiJinAOAS}, the last two authors set out to explore the effects of these data summaries. In \cite{KP:AOAS}, it is shown what information is lost by reducing the data to a simple graph by presenting multi-observation table data summaries, core-decomposition summaries, and hypergraph data summaries, all of which suggest possibly different conclusions than those from the derived simple graphs. 
For example, the authors considered the inner-most clique, that is, the largest completely connected  subgraph,  of the coauthorship graph where there is an edge between two authors if they coauthored at least 4 joint papers. While these authors have many neighbors, i.e. their nodes have a high degree, we argue that degree-based modeling on the simple graph does not capture everything behind the data. Specifically, Figure~\ref{fig:CliqueSecret} shows that the secret behind these cliques is.... a single many-author paper in both cases.

With the issues illustrated in \cite{KP:AOAS} in mind, 
Fienberg and coauthors introduce the $\beta$ model for random hypergraphs in \cite{betaHypergraphs}, which builds upon and generalizes the well-studied $\beta$ model for random graphs. Directly motivated by Fienberg's earlier foundational work, the authors provide two algorithms for fitting the model parameters, an iterative proportional scaling algorithm and a fixed point algorithm.  Furthermore, Fienberg and coauthors prove that both algorithms converge if the maximum likelihood estimator (MLE) exists, and they provide algorithmic and geometric ways of dealing the issue of MLE existence---one of Fienberg's favorite problems. 
\begin{figure}[h]
	\includegraphics[scale=.5]{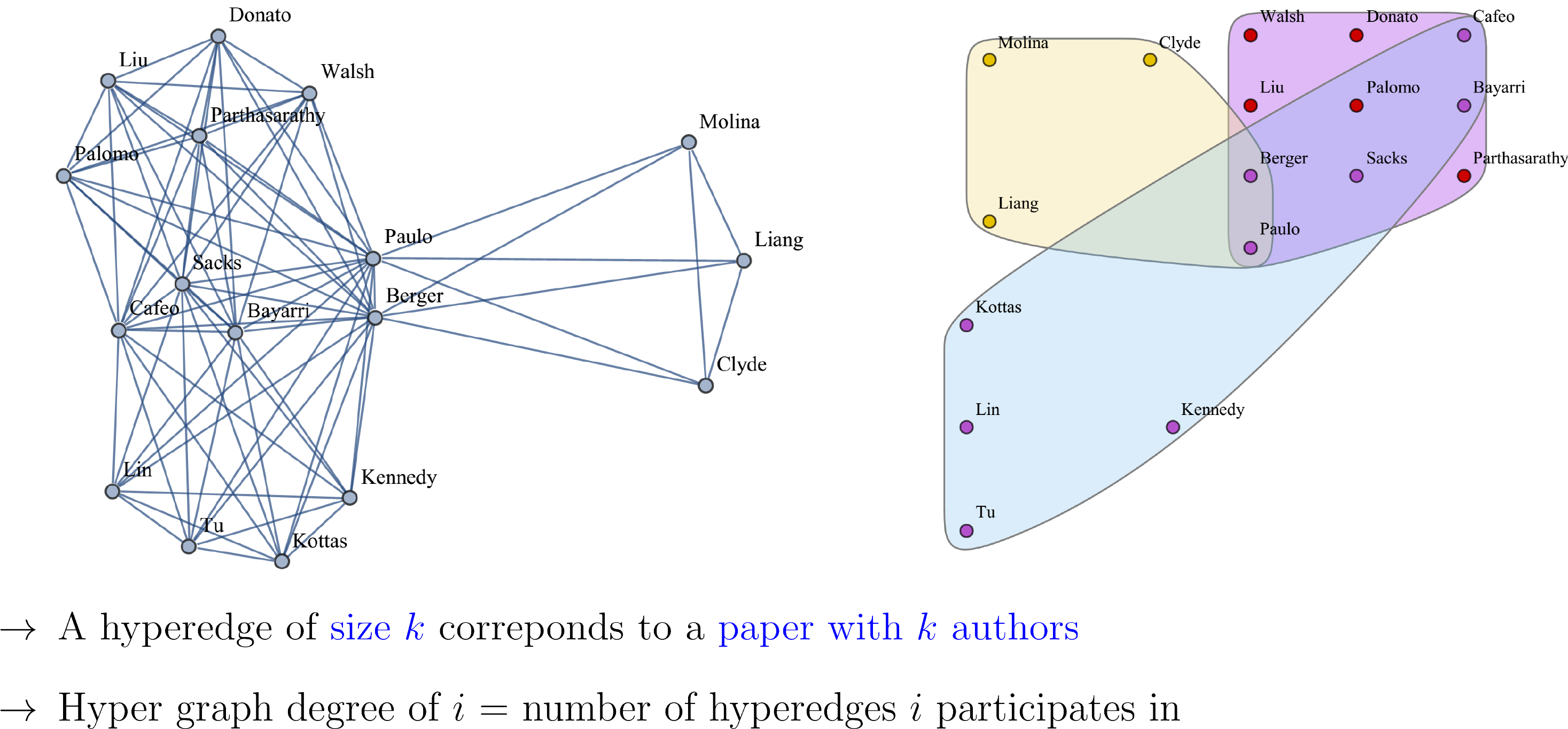}
	\caption{The graph and the hypergraph representing \emph{the same} co-authorship data. In the graph on the left, it is not clear at all that the data corersponds to exactly 3 published papers, for example, which is clear in the hypergraph on the right. 
	 Graphs in the figure adapted from \cite{KP:AOAS}.} 
\label{fig:hypergraphVSgraph}
\end{figure}

\section{Closing remarks} Fienberg always used to say how problems never go away, one just sees them under a new light. In this survey of Fienberg's work connecting categorical data analysis and algebraic statistics to network science, we hope we illustrated, in essence, this sentiment of continual discovery and rediscovery. 

  
\bibliographystyle{alpha}
\bibliography{AlgStatAndNtwks,refAOASvishesh,GoFandSBM,ContTablesSteve}

\end{document}